\documentclass[journal]{rmaa}

\usepackage{paralist}

\usepackage{psfrag,color}

\usepackage[latin1]{inputenc}

\title{Determination of Stellar Atmospheric Parameters for a sample of the post-AGB stars} 

\author{
R. E. Molina,\altaffilmark{1} 
}

\altaffiltext{1}{Laboratorio de Investigaci\'on en F\'{i}sica Aplicada y Computacional,
  Universidad Nacional Experimental del T\'achira, CP 5001, Venezuela.}

\shortauthor{Molina, R.E.}
\shorttitle{T$_\mathrm{eff}$, log\,$g$, [Fe/H] in post-AGB stars}

\fulladdresses{
\item R. E. Molina: Laboratorio de Investigaci\'on en F\'{i}sica Aplicada y
Computacional, Universidad Nacional Experimental del T\'achira,  Venezuela,
(rmolina@unet.edu.ve).}
\listofauthors{R. E. Molina }
\abstract{We report for the first time the stellar atmospheric parameters for a
a set of post-AGB stars classified by Su\'arez et al. (2006). The stellar
spectra were taken from optical region, with low-resolution and have different
spectral ranges. We select a sample of 70 objects with A--K spectral types and
luminosities I and Ie. The large majority of these objects have been scarcely studied
and are located toward the galactic south pole
region. We employ a set of empirical relationships that use pseudo-equivalent
widths like spectral feature to estimate effective temperature, surface
gravity and metallicity. The criteria chosen for selection of absorption are
similar to employed by MK classification system. }

\resumen{Reportamos por primera vez los par\'ametros atmosf\'ericos estelares para un 
conjunto de estrellas clasificadas como post-AGB por Su\'arez et al. (2006). 
Los espectros estelares empleados para el estudio fueron tomados en la
regi\'on \'optica, con baja resoluci\'on y poseen diferentes rangos espectrales. 
Seleccionamos una muestra de 70 objetos con tipos espectrales que abarcan
un rango entre A--K y clases de luminosidades I y Ie. La mayor\'ia de las 
estrellas han sido poco estudiadas y se encuentran ubicadas hacia la regi\'on 
del polo sur gal\'actico. Se emplean un conjunto de calibraciones emp\'iricas
que utilizan los pseudo anchos equivalentes como caracter\'istica espectral
para estimar la temperatura efectiva, la gravedad superficial y la metalicidad.
Los criterios tomados para la selecci\'on de las l\'ineas de absorci\'on son
similares a los empleados por el sistema MK.}

\addkeyword{Stars: equivalent widths}
\addkeyword{Stars: stellar atmospheric parameters}
\addkeyword{Stars: post-AGB}

\begin{document}
% Typeset article header
\maketitle

\section{Introduction}
\label{sec:introd}

When performing a detailed analysis of the chemical abundances, it is essential 
to estimated as accuratelly as possible the relevant physical parameters that
will lead the choise of the proper atmospheric models, i.e. effective temperature, 
surface gravity, and micro-turbulence velocity. This can be archieved from a
variety of photometric (e.g. Arellano Ferro, Mendoza \& Eugenio 1993; Schuster 
et al.\@ 1996; Alonso et al.\@ 1999; Mauro et al.\@ 2013) and spectroscopic
methods (e.g. Gray et al.\@ 2001; Giridhar \& Goswami 2002; Molina \& Stock 2004; 
Soubiran et al.\@ 2010; Wu et al.\@ 2011, Chen et al.\@ 2015; Teixeira et al.\@ 2016). 

Stellar atmosphere is characterized mainly by T$_\mathrm{eff}$, log\,$g$, $\xi_{t}$ 
and [Fe/H], and the knowledge of these parameters is crucial in many research areas
related to the stellar and galaxy physics.

The traditional spectroscopic method to initially derive the effective temperature and
gravity is via the ionization equilibrium of a well represented specie, such as that
of \ion{Fe}{i} and \ion{Fe}{ii} or \ion{Ti}{i} and \ion{Ti}{ii} and a
set of stellar models such as \texttt{ODFNEW-ATLAS9} (Castelli \& Kurucz 2003)
and \texttt{MARCS} (Gustafsson et al.\@ 2008).

Empirical calibrations to estimate the stellar parameters employ, besides equivalent widths,
other quantifiable spectroscopic features such as the central residual intensities
 and pseudocontinuum peaks (Rose 1984), relative depth ratios (Kovtyukh et al.\@ 2003) and
photometric bandheads (\'Arnad\'ottir et al.\@ 2010).
It is important to note that the stellar parameters derived from empirical calibrations
have been one of the main sources of information for the selection and validation of the
stellar model for any object under study.

The large data bases that have become available over the last decade, e.g. \texttt{RAVE} 
(Zwitter et al.\@ 2008), \texttt{APOGEE} (Allende-Prieto et al.\@ 2008), \texttt{LAMOST} 
(Zhao et al.\@ 2012), to name a few, require automated 
processing methods that allow the characterization of high volumes of information 
(stellar spectra) in relatively short time (Graff et al.\@ 2013; Bellinger et al.\@ 2016; 
Dafonte et al.\@ 2016; Damiani et al.\@ 2016; Ren et al.\@ 2016).

Recently, automatic or semiautomatic methods for 
determining equivalent widths and stellar parameters have been developed, such as
\texttt{ROBOSPECT} (Waters \& Hollek 2013), \texttt{GALA} (Mucciarelli 2013), \texttt{FAMA} 
(Magrini et al.\@ 2013), \texttt{ISpec} (Blanco-Cuaresma et al.\@ 2014) and \texttt{ARES+MOOG} 
(Sousa 2014). These methods have been calibrated for a wide range of stars in different 
evolutionary stages from dwarf stars to giant stars. Stellar parameters and elemental 
abundances estimated via these automated methods show some degree of reliability 
and efficiency (Teixeira et al.\@ 2016). 
However, these methods have not been tested for highly evolved objects such as 
post-AGB stars (hereinafter PAGB) due to the peculiarities of their spectra, e.g. complex
emission and absoption profiles, profiles of strong absorption distorted by emission and
splitting, and metal emission features (Klochkova 2014).

This paper aims to estimate T$_\mathrm{eff}$, log\,$g$ and [Fe/H] for a set of stars PAGB
via empirical calibrations of equivalent widths of selected features. Such empirical
calibrations to determine the stellar parameters in PAGB stars are scarces (e.g. 
Arellano Ferro 2010; Molina 2012) as they are sensitive to the fact that these stars may
be variables and their extinction, which is commonly a combination of interstellar and
circumstellar, significantly affect the estimations temperature, gravity and distance
of the central star.

This paper are organized as follows: \S~\ref{sec:sample} describes the selection of the
sample stars and how equivalent widths were determined.
\S~\ref{sec:param} shows the spectroscopic calibrations that allow us to derived the stellar
atmospheric parameters. In \S~\ref{sec:results} is dedicated to 
discuss our results, and finally, in \S~\ref{sec:summconc} gives the conclusions of the paper.

\begin{figure}
\centering
\includegraphics[width=7.5cm,height=7.5cm]{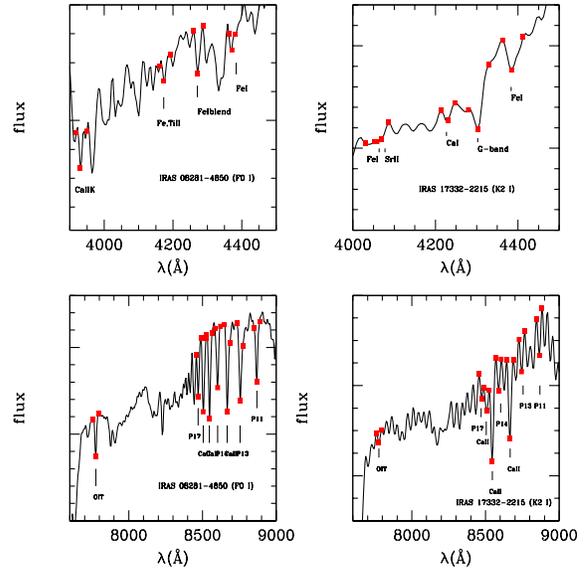}
\caption{Maximum and minimun points (red squares) that allow us to derive the equivalent widths
for IRAS\,08281\,--\,4850 (F0I) and IRAS\,17332\,--\,2215 (K2I) using an automatic code. 
The location of different absorption lines selected are labeled
by continuous lines.}
\label{fig:figure1}
\end{figure}

\begin{table*}
\begin{center}
\begin{minipage}{120mm}
\caption{Equivalent widths for warm stars.}
\label{tab:table1}
\scriptsize{\begin{tabular}{lccccccc}
\noalign{\smallskip}
\toprule
\noalign{\smallskip}
\noalign{\smallskip}
\multicolumn{1}{c}{IRAS}&
\multicolumn{1}{c}{SpT}&
\multicolumn{1}{c}{CaIIK}&
\multicolumn{1}{c}{Fe,TiII}&
\multicolumn{1}{c}{FeI blend}&
\multicolumn{1}{c}{FeI}&
\multicolumn{1}{c}{OI}&
\multicolumn{1}{c}{ref.}\\
\multicolumn{1}{c}{number}&
\multicolumn{1}{c}{}&
\multicolumn{1}{c}{3933}&
\multicolumn{1}{c}{4172-9}&
\multicolumn{1}{c}{4271}&
\multicolumn{1}{c}{4383}&
\multicolumn{1}{c}{7771-5}&
\multicolumn{1}{c}{}\\
\multicolumn{1}{c}{}&
\multicolumn{1}{c}{}&
\multicolumn{1}{c}{(\AA)}&
\multicolumn{1}{c}{(\AA)}&
\multicolumn{1}{c}{(\AA)}&
\multicolumn{1}{c}{(\AA)}&
\multicolumn{1}{c}{(\AA)}&
\multicolumn{1}{c}{}\\
           \noalign{\smallskip}
            \midrule
            \noalign{\smallskip}
02143\,$+$\,5852 &F7Ie&1.01   & 0.18   & 0.32   & 0.05   & $\cdots$ & 01\\
02528\,$-$\,4350 &    &1.26   & 0.19   & 0.13   & $\cdots$ & $\cdots$ &  \\
04296\,$+$\,3429 &F7I&$\cdots$& $\cdots$& $\cdots$& $\cdots$ & $\cdots$ & 01 \\
05341\,$+$\,0852 &F5I&$\cdots$& 1.94   & 0.23   & 0.29   & 1.52 & 01 \\
06530\,$-$\,0213 &F0Iab&$\cdots$& 2.40   & 0.80   & 3.23   & $\cdots$ & 02 \\
07134\,$+$\,1005 &F7Ie&$\cdots$& 1.25   & 0.27   & 0.31   & $\cdots$ & 01 \\
07253\,$-$\,2001 &F2I&$\cdots$& $\cdots$& $\cdots$& 0.27   & $\cdots$ & 01 \\
07430\,$+$\,1115 &G5Ia&$\cdots$& $\cdots$& $\cdots$& 0.99   & $\cdots$ & 04 \\
08005\,$-$\,2356 &F5Ie&4.68   & 1.42   & $\cdots$& 0.73   & 1.27 & 01 \\
08143\,$-$\,4406 &F8I&8.39   & 2.09   & 0.48   & 0.99   & $\cdots$ & 05\\
08187\,$-$\,1905 &F6Ib/II&$\cdots$& $\cdots$& $\cdots$& 0.24   & $\cdots$ & 03 \\
08213\,$-$\,3857 &F2Ie&$\cdots$& 0.74   & 0.19   & 0.22   & $\cdots$ & 01\\
08281\,$-$\,4850 &F0I&1.22   & 0.94   & 2.74   & 1.65   & 1.90 & 01\\
10215\,$-$\,5916 &A7Ie&11.32 & 4.22 & 2.37 & 3.12 & 1.77 & 01\\
10256\,$-$\,5628 &F5I&12.72  & 1.38   & 1.75   & 0.42   & 2.17 & 01 \\
11201\,$-$\,6545 &A3Ie&$\cdots$& $\cdots$& $\cdots$& 0.48   & $\cdots$ & 01\\
11387\,$-$\,6113 &A3Ie&$\cdots$& 0.74   & 0.30   & 0.44   & $\cdots$ & 01\\
12067\,$-$\,4505 &F6I&$\cdots$& $\cdots$& $\cdots$& 0.05   & $\cdots$ & 06\\
14325\,$-$\,6428 &A8I&6.18   & 1.10   & 0.24   & 0.96   & $\cdots$ & 07\\
14429\,$-$\,4539 &A1I&0.91   & 1.05   & 0.23   & 0.47   & 1.85 & 02\\
14482\,$-$\,5725 &A2I&$\cdots$& $\cdots$& $\cdots$& 1.13   & $\cdots$ & 01\\
14488\,$-$\,5405 &A0I&$\cdots$& $\cdots$& $\cdots$& 0.02   & $\cdots$ & 01\\
15039\,$-$\,4806 &A5Iab&0.75   & 0.50   & 0.12   & 0.21   & 1.64 & 08\\
15310\,$-$\,6149 &A7I&$\cdots$& $\cdots$& $\cdots$& 0.09   & $\cdots$ & 01\\
15482\,$-$\,5741 &F7I&$\cdots$& 1.88   & 0.33   & 0.22   & $\cdots$ & 01\\
16283\,$-$\,4424 &A2Ie&4.50   & 0.25   & 3.00   & 1.58   & 2.38 & 01\\
17106\,$-$\,3046 &F5I&9.62   & 2.68   & 1.25   & 0.58   & 2.10 & 01\\
17208\,$-$\,3859 &A2I&$\cdots$& 0.57   & 0.23   & 0.82   & $\cdots$ & 01\\
17245\,$-$\,3951 &F6I&9.13   & 2.00   & 1.49   & 2.13   & 2.02 & 01\\
17287\,$-$\,3443 &   & 0.96  & 0.10   & 0.15   & 0.14   & 0.74 &   \\
17310\,$-$\,3432 &A2I&$\cdots$& 0.36   & 0.22   & $\cdots$&  1.30 & 01\\
17376\,$-$\,2040 &F6I&$\cdots$& $\cdots$& $\cdots$& $\cdots$& $\cdots$ & 01\\
17436\,$+$\,5003 &F3Ib&6.76   & 1.88   & 1.14   & 1.07   & $\cdots$ & 09\\
17441\,$-$\,2411 &F4I&7.38   & 1.16   & 0.99   & 0.52   & 2.06  & 01\\
17488\,$-$\,1741 &F7I&$\cdots$& $\cdots$& $\cdots$& $\cdots$& $\cdots$ & 01\\
17576\,$-$\,2653 &A7I&5.13   & 0.85   & 0.02   & 0.58   & 2.12 & 01\\
17579\,$-$\,3121 &F4I&$\cdots$& 2.42   & 0.24   & 0.98   & $\cdots$ & 01\\
18025\,$-$\,3906 &G1I&13.14  & 2.18   & 0.22   & 1.30   & $\cdots$ & 01\\
18044\,$-$\,1303 &F7I&$\cdots$& $\cdots$& $\cdots$& $\cdots$& $\cdots$ & 01\\
19114\,$+$\,0002 &G5Ia&9.86   & 3.90   & 1.82   & 1.59   & $\cdots$ & 10\\
19207\,$+$\,2023 &F6I&$\cdots$& 3.44   & $\cdots$& 3.74   & $\cdots$ & 01\\
19386\,$+$\,0155 &F5I&8.31   & 1.20   & 1.70   & 0.97   & 1.66 & 11\\
19422\,$+$\,1438 &F5I&$\cdots$& $\cdots$& $\cdots$& $\cdots$& $\cdots$ & 01\\
19500\,$-$\,1709 &F0Ie&1.67   & 0.95   & 0.44   & 0.74   & 1.94 & 01\\
19589\,$+$\,4020 &F5I&$\cdots$& $\cdots$& $\cdots$& $\cdots$& $\cdots$ & 01\\
20160\,$+$\,2734 &F3Ie&4.52   & 1.95   & 1.28   & 1.33   & $\cdots$ & 01\\
20259\,$+$\,4206 &F3I&$\cdots$& $\cdots$& $\cdots$& $\cdots$& $\cdots$ & 01\\
20572\,$+$\,4919 &F3Ie&4.41   & 0.92   & 0.73   & 0.63   & $\cdots$ & 01\\
21289\,$+$\,5815 &A2Ie&0.68   & 1.12   & 1.94   & 0.76   & $\cdots$ & 01\\
22223\,$+$\,4327 &F7I&8.93   & 2.41   & 1.05   & 1.02   & $\cdots$ & 01\\
            \noalign{\smallskip}                                                                                                                                    \bottomrule
            \noalign{\smallskip}  
\end{tabular}}
\end{minipage}
\end{center}
\end{table*}

\begin{table*}
\begin{center}
\begin{minipage}{120mm}
\caption{Equivalent widths for cold stars.}
\label{tab:table2}
\scriptsize{\begin{tabular}{lcccccccc}
\noalign{\smallskip}
\toprule
\noalign{\smallskip}
\noalign{\smallskip}
\multicolumn{1}{c}{IRAS}&
\multicolumn{1}{c}{SpT}&
\multicolumn{1}{c}{FeI}&
\multicolumn{1}{c}{SrII}&
\multicolumn{1}{c}{CaI}&
\multicolumn{1}{c}{G-band}&
\multicolumn{1}{c}{FeI}&
\multicolumn{1}{c}{OI}&
\multicolumn{1}{c}{ref.}\\
\multicolumn{1}{c}{number}&
\multicolumn{1}{c}{}&
\multicolumn{1}{c}{4063}&
\multicolumn{1}{c}{4077}&
\multicolumn{1}{c}{4226}&
\multicolumn{1}{c}{4302}&
\multicolumn{1}{c}{4383}&
\multicolumn{1}{c}{7771-5}&
\multicolumn{1}{c}{}\\
\multicolumn{1}{c}{}&
\multicolumn{1}{c}{}&
\multicolumn{1}{c}{(\AA)}&
\multicolumn{1}{c}{(\AA)}&
\multicolumn{1}{c}{(\AA)}&
\multicolumn{1}{c}{(\AA)}&
\multicolumn{1}{c}{(\AA)}&
\multicolumn{1}{c}{(\AA)}&
\multicolumn{1}{c}{}\\
           \noalign{\smallskip}
            \midrule
            \noalign{\smallskip}
01259\,$+$\,6823&GIab:& 0.76  & 2.95   & 1.34   & 3.60   & 0.48   & $\cdots$ & 12\\
05113\,$+$\,1347&G5I& $\cdots$& $\cdots$& $\cdots$& 1.76  & 1.37   & $\cdots$ & 12\\
05381\,$+$\,1012&G2I& 0.37   & 0.76   & 0.40   & 2.87   & 0.51   & $\cdots$ & 04\\
07331\,$+$\,0021&G5Iab& $\cdots$& 1.70   & 0.77   & 3.60  & 0.88  & $\cdots$ & 10,13\\
07582\,$-$\,4059&G5I& $\cdots$& 1.60   & 0.98   & 5.69   & 2.86   & $\cdots$ & 01\\
10215\,$-$\,5916&   & $\cdots$ & $\cdots$ & 1.53 & 3.03  & 3.12 & 1.77 & 01\\
13203\,$-$\,5917&G2I& $\cdots$& $\cdots$& 5.02   & 6.31   & 4.94  & $\cdots$ & 01\\
13313\,$-$\,5838&K5I& 2.86   & $\cdots$& 2.31   & 4.07   & 1.94   & 0.06 & 01\\
15210\,$-$\,6554&K2I& $\cdots$& 3.28   & 1.54   & 7.65   & 1.87   &  0.98 & 01\\
16494\,$-$\,3930&G2I& $\cdots$& $\cdots$& 1.13   & 2.54   & 0.82  & 1.38 & 01\\
17300\,$-$\,3509&G2I& $\cdots$& 1.18   & 1.08   & 5.21   & 1.30   & $\cdots$ & 01\\
17317\,$-$\,2743&G4I& 2.66   & $\cdots$& 1.59   & 2.06   & 1.68   & 2.07 & 14\\
17332\,$-$\,2215&K2I& $\cdots$& 1.20   & 1.72   & 5.35   & 3.93   &  0.46 & 01\\
17370\,$-$\,3357&G3I& 0.27   & 0.27   & 0.67   & 2.53   & 2.39   &  1.88 & 01\\
17388\,$-$\,2203&G0I& 1.34   & 1.34   & 0.30   & 3.52   & 0.99   &  1.78 & 01\\
18075\,$-$\,0924&G2I& 1.21   & $\cdots$& 0.91   & 1.66   & 1.86   & $\cdots$ & 01\\
18096\,$-$\,3230&G3I& 1.69   & 0.16   & 1.57   & 6.02   & 1.68   & $\cdots$ & 01\\
18582\,$+$\,0001&K2I& $\cdots$& 1.51   & $\cdots$& 4.62   & 2.64  & $\cdots$ & 01\\
19356\,$+$\,0754&K2I& $\cdots$& 2.90   & 1.62   & 6.51   & 2.90   & $\cdots$ & 01\\
19477\,$+$\,2401&G0I& $\cdots$& $\cdots$& $\cdots$& $\cdots$& $\cdots$& $\cdots$ & 14\\
            \noalign{\smallskip}
         \bottomrule
            \noalign{\smallskip}
\end{tabular}}
\end{minipage}
\flushleft{(01) Su\'arez et al.\@ (2006); (02) Hu et al.\@ (1993); (03) Hrivnak et al.\@(1989);
(04) Fujii et al.\@ (2001); (05) Hrivnak \& Bieging (2005); (06) Maas et al.\@ (2002);
(07) Reyniers et al.\@ (2007); (08) Stephenson \& Sanduleak (1971); (09) Min et al.\@ (2013);
(10) Omont et al.\@ (1993); (11) Hrivnak, Lu \& Nault (2015);
(12) Kelly \& Hrivnak (2005); (13) Klochkova (1997); (14) S\'anchez-Contreras et al.\@ (2008)}
\end{center}
\end{table*}

\begin{table*}
\begin{center}
\begin{minipage}{130mm}
\caption{Atmospheric parameters used as calibrators taken from literature.}
\label{tab:table3}
\scriptsize{\begin{tabular}{rlcccc}
\noalign{\smallskip}
\toprule
\noalign{\smallskip}
\multicolumn{1}{c}{IRAS}&
\multicolumn{1}{l}{SpT}&
\multicolumn{1}{c}{T$_\mathrm{eff}$$^\mathrm{ref}$$\pm$$\Delta$T$_\mathrm{eff}$$^\mathrm{ref}$}&
\multicolumn{1}{c}{log\,$g$$^\mathrm{ref}$$\pm$$\Delta$log\,$g$$^\mathrm{ref}$}&
\multicolumn{1}{c}{[Fe/H]$^\mathrm{ref}$$\pm$$\Delta$[Fe/H]$^\mathrm{ref}$}&
\multicolumn{1}{c}{ref.}\\
\multicolumn{1}{c}{number}&
\multicolumn{1}{c}{}&
\multicolumn{1}{c}{(K)}&
\multicolumn{1}{c}{}&
\multicolumn{1}{c}{(dex)}&
\multicolumn{1}{c}{}\\
\noalign{\smallskip}
\midrule
  &  & Warm stars & (6000\,$\leq$\,T$_\mathrm{eff}$\,$\leq$\,8000\,K) &  &  \\
\midrule
\noalign{\smallskip}
15039\,$-$\,4806 & A0I  & 8000$\pm$200 & 1.25$\pm$0.25 & -0.85$\pm$0.10 & 07 \\
14325\,$-$\,6428 & A8I  & 8000$\pm$125 & 1.00$\pm$0.25 & -0.56$\pm$0.16 & 03 \\
19500\,$-$\,1709 & F0Ie & 8000$\pm$125 & 1.00$\pm$0.25 & -0.59$\pm$0.10 & 03 \\
08281\,$-$\,4850 & F0I  & 7875$\pm$125 & 1.25$\pm$0.25 & -0.26$\pm$0.11 & 03 \\
20572\,$+$\,4919  & F3Ie & 7500$\pm$200 & 2.00$\pm$0.50 & -0.01$\pm$0.10 & 13 \\
15482\,$-$\,5741 & F7I  & 7400$\pm$150 & 1.40$\pm$0.20 & -0.47$\pm$0.16 & 08 \\
06530\,$-$\,0213 & F0Iab:& 7375$\pm$125 & 1.25$\pm$0.25 & -0.32$\pm$0.11 & 03 \\
08005\,$-$\,2356 & F5Ie & 7300$\pm$250 & $\cdots$      & $\cdots$       & 06 \\
07134\,$+$\,1005  & F7Ie & 7250$\pm$200 & 0.50$\pm$0.30 & -1.00$\pm$0.20 & 04 \\
08143\,$-$\,4406 & F8I  & 7150$\pm$100 & 1.35$\pm$0.15 & -0.39$\pm$0.12 & 15 \\
17436\,$+$\,5003  & F3Ib & 7065$\pm$125 & 0.91$\pm$0.15 & -0.09$\pm$0.10 & 09 \\
04296\,$+$\,3429  & F7I  & 7000$\pm$250 & 1.00$\pm$0.50 & -0.69$\pm$0.20 & 02 \\
19386\,$+$\,0155  & F5I  & 6800$\pm$100 & 1.40$\pm$0.20 & -1.10$\pm$0.15 & 12 \\
19114\,$+$\,0002  & G5Ia & 6750$\pm$200 & 0.50$\pm$0.25 & -0.45$\pm$0.20 & 11 \\
05341\,$+$\,0852  & F5I  & 6500$\pm$200 & 1.00$\pm$0.50 & -0.72$\pm$0.12 & 04 \\
22223\,$+$\,4327  & F7I  & 6500$\pm$125 & 1.00$\pm$0.25 & -0.30$\pm$0.11 & 03 \\
08187\,$-$\,1905 & F6Ib & 6250$\pm$200 & 0.50$\pm$0.20 & -0.59$\pm$0.15 & 01 \\
18025\,$-$\,3906 & G1I  & 6250$\pm$100 & 0.25$\pm$0.25 & -0.45$\pm$0.16 & 10 \\
20259\,$+$\,4206  & F3I  & 6100$\pm$200 & 2.20$\pm$0.25 & -0.10$\pm$0.15 & 10 \\
12067\,$-$\,4508  & F6I  & 6000$\pm$250 & 1.50$\pm$0.50 & -2.00$\pm$0.12 & 16 \\
07430\,$+$\,1115  & G5Ia & 6000$\pm$125 & 1.00$\pm$0.25 & -0.33$\pm$0.15 & 03 \\
\noalign{\smallskip}
\midrule
  &  & Cool stars & (4500\,$\leq$\,T$_\mathrm{eff}$\,$\leq$\,5500\,K) &  &  \\
\midrule
\noalign{\smallskip}
05113\,+\,1347  & G5I  & 5500$\pm$125 & 0.50$\pm$0.25 & -0.54$\pm$0.17 & 03 \\
05381\,+\,1012  & G2I  & 5200$\pm$100 & 1.00$\pm$0.50 & -0.80$\pm$0.17 & 05 \\
01259\,+\,6823  & GIab:& 5000$\pm$200 & 1.50$\pm$0.25 & -0.60$\pm$0.12 & 01 \\
13313\,$-$\,5838 & K5I  & 4540$\pm$150 & 2.20$\pm$0.30 & -0.09$\pm$0.05 & 14 \\
07331\,+\,0021  & K3/K5I& 4500$\pm$200 & 1.00$\pm$0.25 & -0.16$\pm$0.16 & 01 \\
            \noalign{\smallskip}
            \bottomrule
            \noalign{\smallskip}
\end{tabular}}
\end{minipage}
\flushleft{(01) Rao, Giridhar \& Lambert (2012); (02) Decin et al.\@ (1998); (03) De Smedt et al.\@(2016); 
(04) Reyniers \& van Winckel (2000); (05) Pereira \& Roig (2006); (06) Klochkova (2014);
(07) van Winckel, Oudmaijer \& Trams (1996); (08) Pereira, Gallino \&
Bisterzo (2012); (09) Luck (2014); (10) Molina et al.\@, in preparation; (11) Kipper (2008); (12) Pereira, Lorentz-
Martins \& Manchado (2004); (13) Klochkova et al.\@ (2008); (14) Drake et al.\@ (2012); (15) Reyniers et 
al.\@ (2004); (16) Maas et al.\@ (2002).}
\end{center}
\end{table*}

\section{Stellar sample}
\label{sec:sample}

The sample used in this work was selected from the Su\'arez et al. (2006)'s PAGB list. 
The full sample contains a total of 103 PAGB stars with spectral types ranging from B to M.

\begin{figure}
\centering
\includegraphics[width=7.5cm,height=7.5cm]{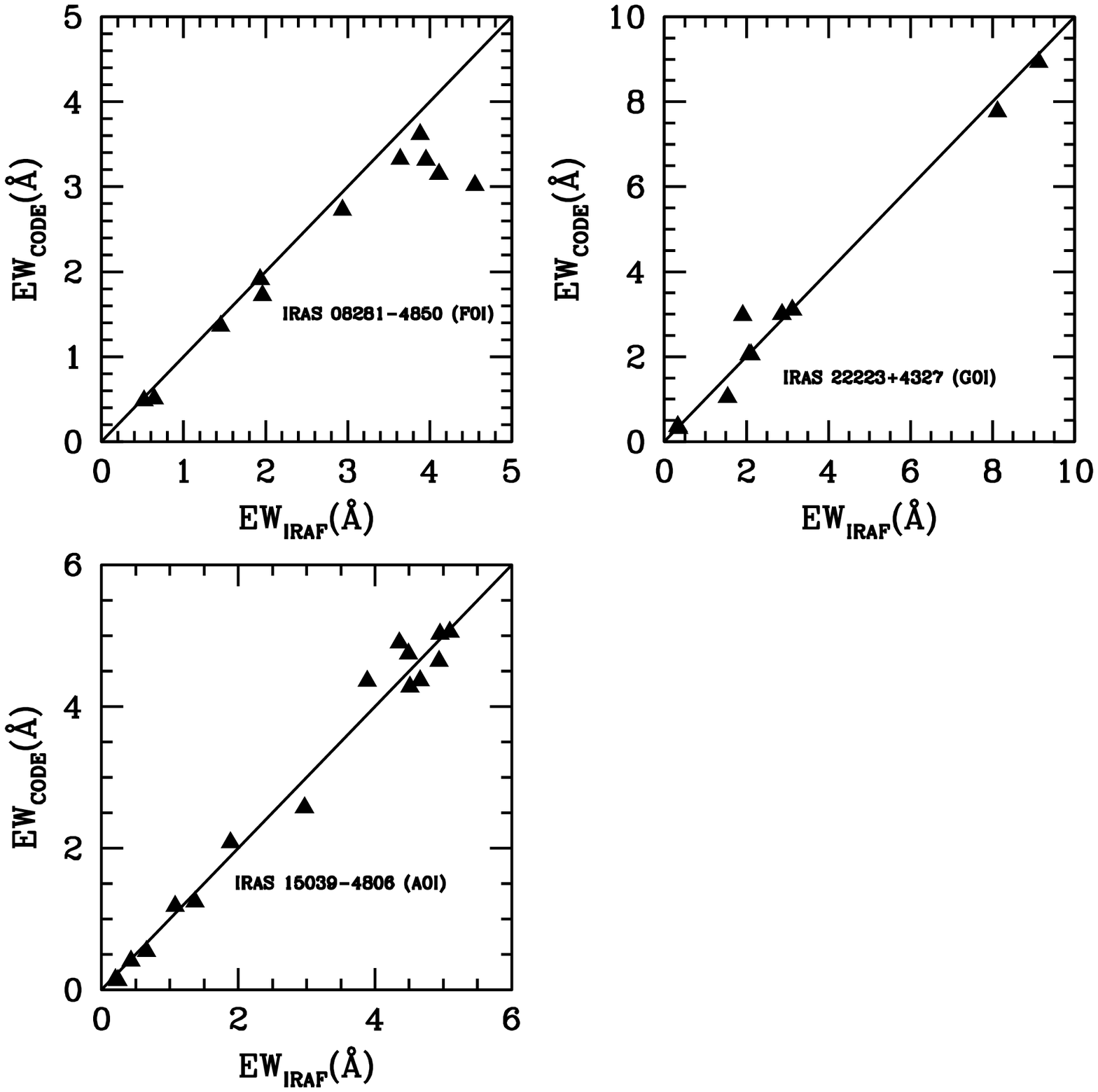}
\caption{Correlation between the equivalent widths measured with the automatic code with those
taken from IRAF code. Three objects IRAS\,01919\,+\,0373,
IRAS\,08281\,-\,4850, IRAS\,22223\,+\,4327 with spectral types A0, F0 and G0 has
been used to compare their results.}
\label{fig:figure2}
\end{figure}

For this work we select a set of 70 PAGB stars with spectral types ranging from A to early-K and 
luminosities classes I and Ie (where ``e'' means emission lines). The selected sample was divided 
in 50 warm (6000\,$\leq$\,T$_\mathrm{eff}$\,$\leq$\,8000\,K) and 20 cold
(4500\,$\leq$\,T$_\mathrm{eff}$\,$\leq$\,5500\,K) PAGB stars.

Subsequently, we identified in the sample those stars that have stellar atmospheric parameters 
T$_\mathrm{eff}$, log\,$g$ and [Fe/H] derived by spectroscopic methods. A total of 21 warm and
5 cold PAGB stars have these parameters determined in the literature with the exception of
IRAS\,08005\,-\,2356, which has only temperature estimation. In warm PAGB stars, some of them
have temperature that are not consistent with their spectral types (i.e. are misclassified). 
Table~\ref{tab:table3} provides the stellar atmospheric parameters used as calibrators collected 
from the literature for the PAGBs in this study.

We choose a total of 9 absorption features which have been widely used as criteria for MK 
spectral classification system. The limitations of some spectra in the spectral range (at 
the blue and near infrared region) make impossible to measure the total equivalent
widths for all objects. In warm-PAGB stars, for example a total of 7 objects do not have 
measures of the equivalent widths and other 8 of them only have a single measure (i.e. the 
\ion{Fe}{i} feature at $\lambda$4383\AA) preventing the estimation of their fundamental parameters.  
In cold-PAGB stars, on the other hand, only one object does not have measures of the equivalent 
width.

\subsection{Determination of equivalent widths}
\label{sec:eqw}

The quantification of equivalent widths was done in an automatic manner. In this sense 
we have developed a code that replaces the true continuum by a pseudo-continuum 
through the interpolation of a straight line that connect the peaks on both sides of an 
absorption line (see Figure~\ref{fig:figure1}). 

The equivalent width is then defined as the effective area occupied between the 
two maximum interpolated $W_{j} = \sum_{j=1}^{n}\frac{(I_{c} - I_{j})}{I_{c}} \Delta\lambda$, 
where $\Delta\lambda$ is a wavelength interval (or its dispersion). Table~\ref{tab:table1} 
and \ref{tab:table2} shows the 
quantified measures of 9 equivalent widths of absorption lines selected in this study.

We can compare the measurements of equivalent widths from the automatic code and those done 
manually with the IRAF code. We use the quantifiable parameters of IRAS\,01919\,+\,0373,
IRAS\,08281\,-\,4850, IRAS\,22223\,+\,4327 with spectral types A0, F0 and G0 respectively.  
From Figure~\ref{fig:figure2} we note that for weaker absorption lines (i.e. with low measures and 
intermediate equivalent widths) their values are in good agreement among themselves, 
while for stronger lines their values show slight systematic differences between them, 
which increase slightly its error. The outliers obtained with this method are usually due to
poorly measured of the equivalent widths caused by a poor maximuum points determination.

\subsection{MK criteria}
\label{sec:mk}

Our atmospheric parameters were estimated from features used by the MK system.

In determining the effective temperature we have used the equivalent widths of the
calcium line \ion{Ca}{ii}K at $\lambda$3933\,\AA\@ (warm stars) and the G-band at $\lambda$4302\,\AA\@ (cool stars).
The \ion{Ca}{ii}K feature grows dramatically in strength of A-type toward to late types ($\sim$F8),
for cooler spectral types their equivalent widths remain flat. Other features as 
\ion{Ca}{i} at $\lambda$4226\,\AA\@ and \ion{Mn}{i} at $\lambda$4030\,\AA\@ blend are not useful to estimating 
the temperature in warm stars.

In the G-type stars, the G-band characteristically dominates over other features. 
This feature increase in strength until about K2 and then decreases in intensity. 
Another feature as \ion{Mg}{i} at $\lambda$5167-72\,\AA\@ triplet shows some sensitivity to temperature 
for cold objects.

For the surface gravity we employ ionized lines like criteria for its determination
($\lambda$4172--79\,\AA\@ and $\lambda$4395-4400\,\AA\@ blends, \ion{Sr}{ii} at $\lambda$4077\,\AA\@ and \ion{Mg}{ii} 
at $\lambda$4481\,\AA\@). It is also possible to use the neutral oxygen (\ion{O}{i} triplet $\lambda$7771-5\,\AA\@)
located in the near IR-region. 
In warm stars,
however, only the $\lambda$4172--79\,\AA\@ blend of \ion{Fe}{ii} and \ion{Ti}{ii} shows sensitivity
to the gravity. While the \ion{Sr}{ii} at $\lambda$4077\,\AA\@ and \ion{Ca}{i} at $\lambda$4226\,\AA\@ lines
show sensitivity to gravity in cold stars.

In order to obtain the stellar metallicity we used only absorption lines of neutral iron,
i.e. \ion{Fe}{i} ($\lambda$4063\,\AA\@), \ion{Fe}{i} ($\lambda$4271\,\AA\@) blend and \ion{Fe}{i} ($\lambda$4383\,\AA\@). We
discard any ionized iron lines because of their expected dependence on log\,$g$. 
In warm stars, we use as metallicity indicator the sum of iron lines \ion{Fe}{i} ($\lambda$4271\,\AA\@ blend +
$\lambda$4383\,\AA\@), while the \ion{Fe}{i} ($\lambda$4063\,\AA\@ and $\lambda$4383\,\AA\@) were employed in cool stars. 

The lines of \ion{Na}{i}D at $\lambda$5889-95\,\AA\@ and \ion{O}{i}T at $\lambda$7771-5\,\AA\@ are used as probable
indicators for the determination of the stellar distance.
The interstellar component of \ion{Na}{i}D lines at $\lambda$5889-95\,\AA\@ show sensitivity to luminosity in
young stars, however, in evolved stars (as PAGB stars) both lines are affected by circumstellar material
and therefore does not show dependence to luminosity. The \ion{O}{i}T lines at $\lambda$7771-5\,\AA\@, on the other 
hand, also show sensitivity to luminosity. In fact, Arellano Ferro et al.\@(2003) found accurate 
spectroscopic calibrations between visual absolute magnitudes and the \ion{O}{i}T lines for a sample
of 27 calibrator stars with spectral types A to G.
  
\begin{figure}
\centering
\includegraphics[width=7.5cm,height=7.5cm]{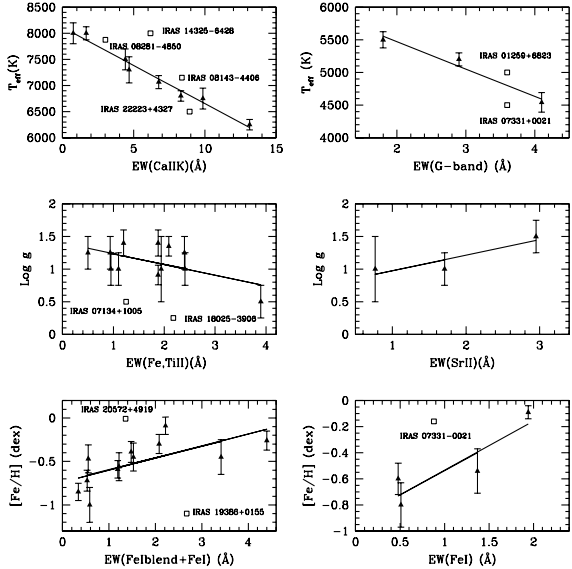}
\caption{Trend of dispersion $\sigma_{EW}$ with the spectral type. We see that dispersion shows
a tendency to increase with the increase of the spectral type in both warm and cold stars.}
\label{fig:figure3}
\end{figure}

Because of the spectral range limitation in the near infrared region, the lines of hydrogen
Paschen series, the oxygen and calcium triplet lines are very rare in the total sample. The
equivalent widths of all features are represented in Table~\ref{tab:table1} and
~\ref{tab:table2}, respectively.

\subsection{Error in the equivalent widths}
\label{sec:error}

An accurate determination of systematic and random errors of the equivalent
widths is not trivial, since these come to be a function of the magnitude, spectral
type, the S/N ratio and the pseudo-continuum position. We also need common stars 
with the same spectral types and several measures of their stellar spectra.
This sample has limitations of objects with the same spectral types and also with 
scarce measurements of equivalent widths, which is impossible to carry out a
reliable statistics.

In this section we can estimate an approximation between the
errors of the equivalent widths of the selected absorption lines and the spectral types.
In this sense, the spectral types were replaced by numerical values in the following sequence:
A0\,=\,30, F0\,=\,40, G0\,=\,50 and K0\,=\,60 respectively. The intermediate values are between 
two successive classes. In view of the difficulties presented by the observational data
(mentioned in the above paragraph), we decided to correlate the equivalent widths determined
through the automatic code with those obtained from the IRAF code for both samples as
shown in Figure~\ref{fig:figure2}.

A dispersion, $\sigma_{EW}$, is obtained for each spectral type (or an average $\sigma_{EW}$, if
the spectral type is repeated), which varies from 0.02\@ to 0.47\@ in the warm stars and
from 0.15\@ to 0.35\@ in the cold stars. In Figure~\ref{fig:figure3}
we observe that the dispersion shows a tendency to increase with the increase of the spectral
type in both warm and cold stars. This dispersion results in an error in the effective 
temperature, $\Delta$T$_\mathrm{eff}$, such that using the equation~\ref{eq1} leads to a
variation between 5\,K to 68\,K and from 63\,K to 122\,K using equations~\ref{eq2} 
and~\ref{eq3} respectively.

With these arguments we can infer that the new spectroscopic calibrations in effective
temperature, gravity and metallicity are not affected by the dispersion in the equivalent withs. 

\subsection{Sample for calibration}
\label{sec:calibra}

Table~\ref{tab:table3} shows the PAGB stars that have been studied and reported in 
the literature. This table contains the number IRAS, spectral type, the stellar atmospheric 
parameters obtained from different sources and their respective references. Their stellar
parameters were obtanied by different authors using spectroscopic methods.
 
We can observe that there are a total of 21 stars considered as warm-PAGB stars and a very small
number of only 5 objects for the cold-PAGB stars.  
In spite of having a small number of stars as calibrators is possible to obtain a rapid and accurate
determination of fundamental parameters (effective temperature, the surface gravity and the 
metallicity) using only suitable spectral criteria, avoiding photometric indices which are often 
distorted by poor known interstellar and circumstellar reddening. 

In the recent past, two papers that involve photometric 
calibrations (Str\"omgren and 2MASS photometry) and that allows to estimate the stellar parameters 
for a group of post-AGB and RV Tauri stars were done by Arellano Ferro et al.\@ (2010) and Molina (2012). 

\begin{table*}
\scriptsize{\begin{center}
\begin{minipage}{90mm}
  \caption{Atmospheric parameters estimated from equivalent widths for warm stars.}
 \label{tab:table4}
\begin{tabular}{lccccr}
\noalign{\smallskip}
\toprule
\noalign{\smallskip}
\noalign{\smallskip}
\multicolumn{1}{c}{IRAS}&
\multicolumn{1}{c}{T$_\mathrm{eff}$$^\mathrm{phot}$}&
\multicolumn{1}{c}{T$_\mathrm{eff}$$^\mathrm{eq1}$}&
\multicolumn{1}{c}{log\,$g$$^\mathrm{phot}$}&
\multicolumn{1}{c}{log\,$g$$^\mathrm{eq4}$}&
\multicolumn{1}{r}{[Fe/H]$^\mathrm{eq7}$}\\
\multicolumn{1}{c}{number}&
\multicolumn{1}{c}{($\pm$ 220K)}&
\multicolumn{1}{c}{($\pm$ 91K)}&
\multicolumn{1}{c}{($\pm$0.27)}&
\multicolumn{1}{c}{($\pm$0.21)}&
\multicolumn{1}{r}{($\pm$0.19)}\\
           \noalign{\smallskip}
            \midrule
            \noalign{\smallskip}
02143\,$+$\,5852  & $\cdots$ & 7967     & $\cdots$ & $\cdots$ & $-$0.68  \\
02528\,$-$\,4350  & $\cdots$ & 7981     & $\cdots$ & $\cdots$ & $-$0.71  \\
07253\,$-$\,2001  & $\cdots$ & 7826     & 1.39     & 1.28     & $-$0.81 \\
08005\,$-$\,2356  & $\cdots$ & $\cdots$ & 1.32     & 1.17     & $-$0.92 \\
08213\,$-$\,3857  & $\cdots$ & 7872     & $\cdots$ & 1.28     & $-$0.67  \\
10215\,$-$\,5916  & $\cdots$ & 6461     & $\cdots$ & $\cdots$ & $\cdots$ \\
10256\,$-$\,5628  & $\cdots$ & 6257     & 0.85     & 1.18     & $-$0.42  \\ 
11201\,$-$\,6545  & $\cdots$ & 7723     & $\cdots$ & 1.26     & $-$0.86 \\
11387\,$-$\,6113  & 6209     & 7707     & 0.75     & 1.28     & $-$0.63  \\
13245\,$-$\,5036\tabnotemark{1}  & 7077     & $\cdots$ & 0.77     & $\cdots$ & $\cdots$ \\
14429\,$-$\,4539  & $\cdots$ & 7981     & 0.95     & 1.23     & $-$0.63  \\
14482\,$-$\,5725  & $\cdots$ & 7402     & $\cdots$ & 1.13     & $-$1.01  \\
14488\,$-$\,5405  & 7578     & 7950     & 0.86     & 1.28     & $-$0.75 \\
15310\,$-$\,6149  & 5787     & 7915     & 0.98     & 1.35     & $-$0.77 \\
16206\,$-$\,5956\tabnotemark{1}  & 7382     & $\cdots$ & 0.86     & $\cdots$ & $\cdots$ \\
16283\,$-$\,4424  & 5699     & 7457     & 0.82     & $\cdots$ & $-$0.09  \\
17106\,$-$\,3046  & $\cdots$ & 6709     & $\cdots$ & 0.97     & $-$0.47  \\
17208\,$-$\,3859  & 5734     & 7856     & 0.96     & 1.31     & $-$0.58  \\
17245\,$-$\,3951  & $\cdots$ & 6781     & 0.91     & 1.08     & $-$0.22  \\
17287\,$-$\,3443  & $\cdots$ & 8024     & $\cdots$ & $\cdots$ & $-$0.69  \\
17310\,$-$\,3432  & $\cdots$ & 7869     & $\cdots$ & $\cdots$ & $\cdots$ \\
17376\,$-$\,2040\tabnotemark{2}  & $\cdots$ & $\cdots$ & $\cdots$ & $\cdots$ & $\cdots$ \\
17441\,$-$\,2411  & 5404     & 7037     & 0.95     & 1.21     & $-$0.52  \\
17488\,$-$\,1741\tabnotemark{2}  & 5860     & $\cdots$ & 0.73     & $\cdots$ & $\cdots$ \\
17576\,$-$\,2653  & 7026     & 7365     & $\cdots$ & 1.26     & $-$0.64  \\
17579\,$-$\,3121  & 5845     & 7790     & 0.78     & 1.01     & $-$0.56  \\
18044\,$-$\,1303\tabnotemark{2}  & $\cdots$ & $\cdots$ & $\cdots$ & $\cdots$ & $\cdots$ \\
19207\,$+$\,2023  & 4785     & 6638     & 0.71     & 0.85     & $\cdots$ \\
19422\,$+$\,1438\tabnotemark{2}  & 6383     & $\cdots$ & 0.72     & $\cdots$ & $\cdots$ \\
19589\,$+$\,4020\tabnotemark{2}  & 5231     & $\cdots$ & 0.78     & $\cdots$ & $\cdots$ \\
20160\,$+$\,2734  & 6168     & 7454     & 0.80     & 1.09     & $-$0.36  \\
21289\,$+$\,5815  & $\cdots$ & 8015     & $\cdots$ & 1.22     & $-$0.35  \\
            \noalign{\smallskip}
            \bottomrule
            \noalign{\smallskip}
\end{tabular}
\end{minipage}
\\
$^{1}$ Emission lines.\\
$^{2}$ Not has measured EWs.
\end{center}}
\end{table*}

\begin{table*}
\begin{center}
\begin{minipage}{120mm}
  \caption{Atmospheric parameters estimated from equivalent widths for cold stars.}
 \label{tab:table5}
\scriptsize{\begin{tabular}{rcccccl}
\noalign{\smallskip}
\toprule
\noalign{\smallskip}
\noalign{\smallskip}
\multicolumn{1}{c}{IRAS}&
\multicolumn{1}{c}{T$_\mathrm{eff}$$^\mathrm{phot}$}&
\multicolumn{1}{c}{T$_\mathrm{eff}$$^\mathrm{eq2}$}&
\multicolumn{1}{c}{T$_\mathrm{eff}$$^\mathrm{eq3}$}&
\multicolumn{1}{c}{log\,$g$$^\mathrm{phot}$}&
\multicolumn{1}{c}{log\,$g$$^\mathrm{eq6}$}&
\multicolumn{1}{r}{[Fe/H]$^\mathrm{eq8}$}\\
\multicolumn{1}{c}{number}&
\multicolumn{1}{c}{($\pm$ 220K)}&
\multicolumn{1}{c}{($\pm$207K)}&
\multicolumn{1}{c}{($\pm$175K)}&
\multicolumn{1}{c}{($\pm$0.27)}&
\multicolumn{1}{c}{($\pm$0.20)}&
\multicolumn{1}{r}{($\pm$0.30)}\\
           \noalign{\smallskip}
            \midrule
            \noalign{\smallskip}
07582\,$-$\,4059 & $\cdots$ & $\cdots$ & 5042     & $\cdots$ & 1.12     & $\cdots$ \\
10215\,$-$\,5916 & $\cdots$ & 5027     & 4852     & $\cdots$ & $\cdots$ & $\cdots$ \\
13203\,$-$\,5917 & 6355     & $\cdots$ & $\cdots$ & $\cdots$ & 1.23     & $\cdots$ \\
15210\,$-$\,6554 & $\cdots$ & $\cdots$ & 4848     & 0.74     & $\cdots$ & $-$0.21 \\
16494\,$-$\,3930 & 6227     & 5232     & 4990     & $\cdots$ & 1.15     & $-$0.61 \\
17300\,$-$\,3509 & $\cdots$ & $\cdots$ & 5007     & 1.10     & 1.02     & $-$0.43 \\
17317\,$-$\,2743 & $\cdots$ & 5432     & 4831     & $\cdots$ & 1.14     & $-$0.28 \\
17332\,$-$\,2215 & $\cdots$ & $\cdots$ & 4786     & $\cdots$ & 1.03     & $\cdots$ \\
17370\,$-$\,3357 & 4869     & 5236     & 5149     & $\cdots$ & $\cdots$ & $-$0.78\tabnotemark{a} \\
17388\,$-$\,2203 & 5267     & 4823     & $\cdots$ & $\cdots$ & 1.06     & $-$0.54 \\
18075\,$-$\,0924 & 5517     & 5599     & 5066     & $\cdots$ & 1.23     & $-$0.21 \\
18096\,$-$\,3230 & $\cdots$ & $\cdots$ & 4838     & 0.75     & $\cdots$ & $-$0.28 \\
18582\,$+$\,0001 & $\cdots$ & $\cdots$ & $\cdots$ & $\cdots$ & 1.10     & $\cdots$ \\
19356\,$+$\,0754 & $\cdots$ & $\cdots$ & 4820     & $\cdots$ & 1.44     & $\cdots$ \\
19477\,$+$\,2401 & $\cdots$ & $\cdots$ & $\cdots$ & $\cdots$ & $\cdots$ \\
            \noalign{\smallskip}
            \bottomrule
            \noalign{\smallskip}
\end{tabular}}
\end{minipage}
\end{center}
\end{table*}

\subsection{Polynomial's fitting}
\label{sec:polyn}

The stellar atmospheric parameters can be determined by fitting a series of polinomials
whose independent variables are equivalent widths. Our goal is to analyze 
the actual dependence of the stellar parameters with respect to one or two quantifiable features. 
The mathematical representation of the polynomial, in general, has the form
\begin{eqnarray}\nonumber
V&=&a_{00}x+a_{01}y+a_{02}xy,
\end{eqnarray}

\noindent where $V$ is any of the three stellar parameters (T$_\mathrm{eff}$, log\,$g$ and [Fe/H]),
a$_{ij}$ are the coefficients to determine and $x$ and $y$ are the independent variables. 
When the number of independent variables is greater than one we used the method adopted
by Stock \& Stock (1999). This method developed a quantitative method to obtain
stellar physical parameters such as absolute magnitude, intrinsic colour, and a
metallicity index using the equivalent
widths of absortion features in stellar spectra by means of polynomials and a
consistent algorithm (Molina \& Stock 2004).

In order to determine the best coefficients we employ an algorithm based on least squares. This
algorithm performs an initial fitting and removes those values of residuals greater than 2-$\sigma$.
The error of each coefficient is obtained from
\begin{eqnarray}\nonumber
\sigma_{a_{ij}}^{2}&=&\sigma^{2} S(i,j),
\end{eqnarray}
 
\noindent where $S(i,j)$ the diagonal matriz and $\sigma^{2}$ is the mean square error. 

\section{Stellar atmospheric parameters}
\label{sec:param}

The main objective of this work is to build a set of spectroscopy calibrations to derive T$_\mathrm{eff}$,
log\,$g$ and [Fe/H] for PAGB stars. We employ the data contained in Tables~\ref{tab:table1},
\ref{tab:table2} and \@\ref{tab:table3}.
In this section we will show the best fits when comparing the
equivalent widths with the stellar atmospheric parameters taken from literature.

\begin{figure}
\centering
\includegraphics[width=7.5cm,height=7.5cm]{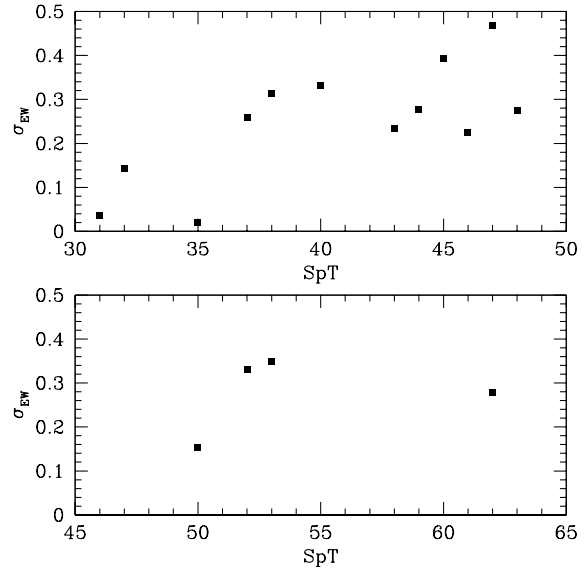}
\caption{Relation between the stellar atmospheric parameters as a function of
equivalent widths for warm stars (left panel) and cold stars (right panel).
Left panel. The empty squares represent those stars that left out of the best fit.}
\label{fig:figure4}
\end{figure}

\subsection{T$_\mathrm{eff}$'s calibration}
\label{sec:temp}  

For the determination of effective temperature in warm-PAGB stars we use equivalent widths of 
the \ion{Ca}{ii}K at $\lambda$3933\,\AA\@. This line has been considered in the MK system as an indicator 
of temperature in warm stars (Gray \& Corbally 2008). 
Particularly, for stars with temperature between (6000\,$\leq$\,T$_\mathrm{eff}$\,$\leq$\,8000\,K),
the equivalent widths show sensitivity to effective temperature. A code based on least squares that 
relates equivalent widths and the effective temperature taken from literature 
(T$_\mathrm{eff}$$^\mathrm{ref}$) leads to the following relationship
\begin{eqnarray}\label{eq1}
T_\mathrm{eff}\!\!\!&\!\!\!=\!\!\!&\!\!\!(8114\pm65)-(146\pm24)(\ion{Ca}{ii}K), 
\end{eqnarray}

\noindent where this calibration is valid for a range in the equivalent widths between
0.76\,$\leq$\,T$_\mathrm{eff}$\,$\leq$ 13.15\,\AA\@. The standard deviation derived from the 
equation~(\ref{eq1}) is $\pm$91\,K. Four stars are left out of the fit, i.e. IRAS\,08143\,--\,4406,
IRAS\,08281\,--\,4850, IRAS\,14325\,--\,6428 and IRAS\,22223\,+\,4327 respectively. 
The stellar temperature estimated by De Smedt et al. (2016) for IRAS\,08281\,--\,4850, 
IRAS\,14325\,--\,6428 and IRAS\,22223\,+\,4327 are 7875\,K, 8000\,K and 6500\,K and Reyniers et al. 
(2004) for IRAS\,08143\,--\,4406 is 7150\,K, while the fit of eq.~(\ref{eq1}) leads to values of
7674\,K, 7211\,K, 6809\,K and 6856\,K respectively.

In late-PAGB stars (4500\,$\leq$\,T$_\mathrm{eff}$\,$\leq$\,5500\,K), on the other hand, 
is possible 
to determine the effective temperature from the G-band at $\lambda$4302\,\AA\@. In spite of only 5 stars 
are present in the fit, is possible therefore to determine the effective temperature applying a 
linear fit
\begin{eqnarray}\label{eq2}
T_\mathrm{eff}\!\!\!&\!\!\!=\!\!\!&\!\!\!(6291\pm246)-(417\pm81)(Gband).
\end{eqnarray}

The equation~(\ref{eq2}) is valid for a range in the equivalent widths between
1.76\,$\leq$\,T$_\mathrm{eff}$\,$\leq$ 4.08\,\AA\@ and where the standard deviation reached is
$\pm$207\,K. Two objects are left out of this relationship, IRAS\,01259\,+\,6823 (5000\,K), 
IRAS\,22223\,+\,4327 (4500\,K) and the fit for both objects reaches the same temperature
value of 4788\,K\@.

The stellar temperature for identified cold PAGB stars can be increased by using the resonance 
\ion{Ca}{i} ($\lambda$4226\,\AA\@) line. This line is sensitive to temperature, since it grows gradually
from the G-type to the early K-type stars being stronger in those stars with mid-K. A lineal relationship
can be obtained by adjusting the temperature and the \ion{Ca}{i} equivalent widths for four
calibrating stars, this is 
\begin{eqnarray}\label{eq3}
T_\mathrm{eff}\!\!\!&\!\!\!=\!\!\!&\!\!\!(5381\pm118)-(436\pm76)(CaI),
\end{eqnarray}

\noindent where its standard deviation reaches a value of $\pm$175\,K and the validation range for 
equivalent widths can be found between 0.40\,\AA\@ and 2.31\,\AA\@ and the temperature 
between $\lambda$4550\,\AA\@ and $\lambda$5200\,\AA\@. 
The results of effective temperature estimated by equations~(\ref{eq1}),~(\ref{eq2}) and~(\ref{eq3}) are
in the third and fourth column of Tables~\ref{tab:table4} and ~\ref{tab:table5}. In the top of 
Figure~\ref{fig:figure4}, we note the dependence of the \ion{Ca}{ii}K-line and the G-band 
with the effective temperature (see left and right panels). 

\subsection{Log\,$g$'s calibration}
\label{sec:grav}

In warm stars, we can estimate the surface gravity using the Fe,TiII blend at $\lambda$4172-9\,\AA\@. 
This blend is constituted mainly by ionized lines of Fe and Ti and has been considered as
indicator as luminosity in A-F type stars. A lineal fit leads to the
following relationship
\begin{eqnarray}\label{eq4}
log\,g\!\!\!&\!\!\!=\!\!\!&\!\!\!(1.40\pm0.14)-(0.16\pm0.07)(FeTiII).
\end{eqnarray}

The range of validation of this calibration in surface gravity covers 0.50\,$\leq$log\,$g$\,$\leq$\,1.40, 
while the equivalent widths of the ionized line vary between 
0.50\,$\leq$\,\ion{FeTi}{ii}\,$\leq$\,3.90\,\AA\@.
The standard desviation leads to a value of $\sigma$=$\pm$0.21.
Two stars fall out of the fit of eq~(\ref{eq4}), i.e. IRAS\,07134\,+\,1005 and IRAS\,18025\,--\,3906. 
According to spectral types (or effective temperature), IRAS\,07134\,+\,1005 and IRAS\,18025\,--\,3906 
it would be expected that their equivalent widths were slightly greater than 4\,\AA\@.

We can extend the range in the surface gravity at higher values using the \ion{O}{i} triplet
lines. Due to the limitations of the spectral range to the near infrared region, the number of 
\ion{O}{i} triplet lines are very scarce. Even though their values are not report in 
Table~\ref{tab:table4}, and we will only show the functional relationship
\begin{eqnarray}\label{eq5}
log\,g\!\!\!&\!\!\!=\!\!\!&\!\!\!(2.20\pm0.20)-(0.58\pm0.13)(OIT),
\end{eqnarray}

\noindent where the range on gravity vary from 1.00 to 2.20 and their equivalent widths between
0.07 to 1.95 \AA\@ respectively. The standard desviation leads a value of $\sigma$=$\pm$0.35.

In cold stars, the surface gravity is estimated using the \ion{Sr}{ii}-line. This line has
been considered as the principal luminosity discriminator for cool stars in MK classification.
Unfortunately the functional relationship is built with only 3 stars and this has the following
form
\begin{eqnarray}\label{eq6}
log\,g\!\!\!&\!\!\!=\!\!\!&\!\!\!(0.74\pm0.23)+(0.24\pm0.11)(SrII).
\end{eqnarray}

The range of validation of this calibration in surface gravity covers 1.00\,$\leq$log\,$g$\,$\leq$\,1.50,
while the equivalent widths of the ionized line vary between
0.77\,$\leq$\,\ion{Sr}{ii}\,$\leq$\,2.95 \AA\@.
The standard desviation leads to a value of $\sigma$=$\pm$0.20.
We can also estimate the gravity for additional cold PAGB stars 13203$-$5917, 16494$-$3930,
17317$-$2743 and 17388$-$2203 when recovering the equivalent widths of the \ion{Sr}{ii} line from
the \ion{Mg}{ii} line. An error of $\pm$0.25 is introduced when making this estimation.

The results of surface gravity estimated by equations~(\ref{eq4}) and ~(\ref{eq6}) are
in the fifth and sixth column of Tables~\ref{tab:table4} and ~\ref{tab:table5}. In the middle of
Figure~\ref{fig:figure4}, we see the dependence of the \ion{Fe,Ti}{ii} blend and the \ion{Sr}{ii}
with regard to surface gravity (see left and right panels).

\subsection{[Fe/H]'s calibration}
\label{sec:metal}

For the calibration of metallicity we used only neutral Fe lines. In warm stars, we use
the sum of \ion{Fe}{i} ($\lambda$4271\,\AA\@ + $\lambda$4383\,\AA\@).
The best fitting that recovers the metallicity is generated by a polinomial that have the form
\begin{eqnarray}\label{eq7}
[Fe/H]\!\!\!&\!\!\!=\!\!\!&\!\!\!-(0.73\pm0.10)+(0.14\pm0.05) \nonumber \\
& & {} (FeI blend+FeI).
\end{eqnarray}

The range of validation of this calibration on metallicity covers $-$0.09\,$\leq$\,[Fe/H]\,$\leq$\,$-$1.00\@ dex,
while the equivalent widths of Fe lines vary between 0.34\,$\leq$\,\ion{Fe}{i}\,$\leq$\,4.40 \AA\@.
The standard desviation for this relationship is $\pm$0.19\,dex. Two outliers are present in this fitting;
IRAS\,19386\,+\,0155 to very low metallicity ($-$1.00 dex) and IRAS\,20572\,+\,4919 to solar metallicity 
($-$0.01 dex) respectively.

In cold stars, we employ the \ion{Fe}{i} lines at $\lambda$4063\,\AA\@ and $\lambda$4383\,\AA\@. Of the 17 cold-PAGB stars
only 5 objects have identified stellar parameters. For the \ion{Fe}{i} ($\lambda$4353\,\AA) line the five objects are
available for the calibration. The best fitting that recovers
the metallicity within a range of $-$0.09\,$\leq$\,[Fe/H]\,$\leq$\,$-$0.80 dex, involves a lineal polynomial
for \ion{Fe}{i} line at $\lambda$4383\,\AA\@, that is
\begin{eqnarray}\label{eq8}
[Fe/H]\!\!\!&\!\!\!=\!\!\!&\!\!\!-(0.92\pm0.16)+(0.38\pm0.13)(FeI),
\end{eqnarray}

\noindent where the range of equivalent widths vary between 0.48 to 1.94 \AA\@ and the standard 
desviation leads a value of $\sigma$=$\pm$0.30 dex.

On the contrary, the best fitting for \ion{Fe}{i} line at $\lambda$4063\,\AA\@ has the form
\begin{eqnarray}\label{eq9}
[Fe/H]\!\!\!&\!\!\!=\!\!\!&\!\!\!-\!(0.85\pm0.18)\!+\!(0.26\pm0.11)\!(FeI).
\end{eqnarray}

The range of equivalent widths vary between 0.48 to 2.86 \AA\@ and the standard
desviation leads a value of $\sigma$=$\pm$0.30 dex.

The results of metallicity estimated by equations~(\ref{eq7}) and ~(\ref{eq8}) are
in the sixth and seventh column of Tables~\ref{tab:table4} and ~\ref{tab:table5}. In the bottom of
Figure~\ref{fig:figure4} we observe the dependence of the \ion{Fe}{i} lines with regard to metallicity 
(see left and right panels).

\section {Results and discussion}
\label{sec:results}

The results of the stellar parameters (columns 3, 5 and 6) 
for the sample studied are shown in Tables~\ref{tab:table4}, \ref{tab:table5} respectively. 
In general, the limitation in the spectral range and the low number of objects with identified
stellar parameters lead to the fact that spectroscopic calibrations can not be applied 
individually to the total sample studied.

For the warm-PAGB stars, we observe that the \ion{Ca}{ii}K line show a strong dependence on the
effective temperature (see Fig.~\ref{fig:figure4}). However, the equivalent widths have 
been measured only for 9 objects out of a total of 29 identified. In order to expand the number
of objects with the new values of T$_\mathrm{eff}$, we estimate the equivalent widths
of the \ion{Ca}{ii}K line from Fe,Ti\,II ($\lambda$4172-9\,\AA) blend and \ion{Fe}{i} ($\lambda$4383\,\AA). 

Clearly this procedure introduces an uncertainty of $\pm$220\,K to the temperature of the additional PAGB
stars, i.e. 07253$-$2001, 08213$-$3857, 11201$-$6545, 11387$-$6113, 14482$-$5725, 14488$-$5405,
15310$-$6149, 17208$-$3859, 17310$-$3432, 17579$-$3121 and 19207$+$2023 respectively. 
A similar procedure has been applied to surface gravity and metallicity in 
order to add new values for those objects not studied.

\begin{figure}
\centering
\includegraphics[width=7.5cm,height=7.5cm]{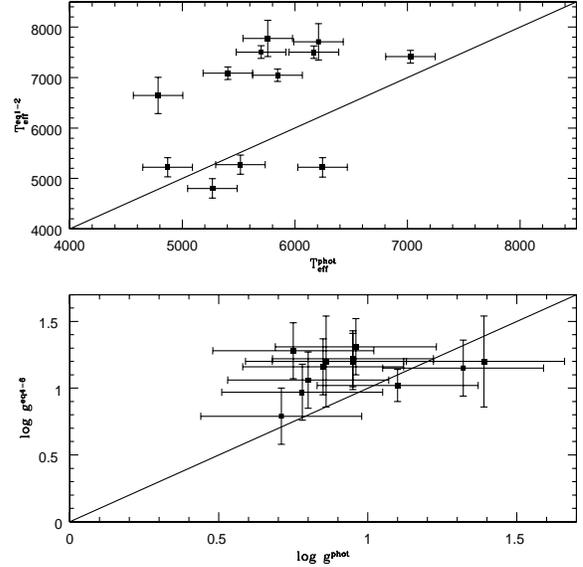}
\caption{Comparation between effective temperature and surface gravity obtained from 2MASS 
photometry by Molina (2012) and effective temperature and surface gravity estimated 
in this work (see upper and bottom panels). 
The solid stright line in each panel denotes perfect agreement between the sets of data.}
\label{fig:figure5}
\end{figure}

For surface gravity the equivalent widths of the Fe,Ti II ($\lambda$4172-9\,\AA) blend is
derived from the \ion{Mg}{ii} ($\lambda$4481\,\AA) line and 4 PAGB stars (07253$-$2001, 11201$-$6545, 
14482$-$5725 and 14488$-$5405) were added with an uncertainty of $\pm$0.30.
The \ion{O}{i} ($\lambda$7771-5\,\AA) triplet line can also be used to determine the surface
gravity of those stars with 1.0\,$\leq$\,log\,$g$\,$\leq$\,2.2 respectively. According to
MK classification system the \ion{O}{i} triplet is sensitive to luminosity (or gravity).

In metallicity the neutral iron blend of \ion{Fe}{i} ($\lambda$4271\,\AA) is determined from 
\ion{Fe}{i} ($\lambda$4383\,\AA) line. An uncertainty of $\pm$0.31 dex is estimated for additional 
PAGB stars; 07253$-$2001, 08005$-$2356, 11201$-$6545, 14482$-$5725, 14488$-$5405 and
15310$-$6149 respectively.

In cold-PAGB stars, however, the G-band and the \ion{Fe}{i} (4383\,\AA) line have measures of
equivalent widths for most objects, except the \ion{Sr}{ii} (4077\,\AA) and \ion{Ca}{i} 
($\lambda$4226\,\AA) line
that is present only for 12 and 17 objects. Unfortunately, the number of objects with identified
stellar parameters is very scarse, which means that the calibrations made are few unreliable.
The results in the metallicity that have a subindex ``a'' represent the values obtained
from eq.~\ref{eq9}. 

We can compare our results in T$_\mathrm{eff}$ and log\,$g$ with a source whose values come
from photometric calibrations for PAGB and RV Tauri stars (Molina 2012).
The values of T$_\mathrm{eff}$ and log\,$g$ determined from the photometric calibrations
are found in the second and fourth columns of Tables~\ref{tab:table4} and \ref{tab:table5},
respectively. In the upper and bottom panels of Figure~\ref{fig:figure5} we can see the comparison
between the spectroscopic and photometric calibrations.

From the Figure~\ref{fig:figure5} we can observe that the T$_\mathrm{eff}^{spec}$ and log\,$g^{spec}$ 
obtained 
spectroscopically from eq.~(\ref{eq1}) and (\ref{eq3}) (warm stars) and from eq~(\ref{eq2}) 
and (\ref{eq5}) (cold stars) are slightly higher than T$_\mathrm{eff}^{phot}$ and log\,$g^{phot}$
obtained photometrically from Molina's calibrations. PAGB stars with temperature close to
5000\,K seem to be adjusted satisfactorily but at a higher temperature the dispersion increase. 
In surface gravity, on the other hand, the spectroscopic values seem to show agreement within 
their uncertainties with photometric values. These results indicate that the interstellar and
circumstellar reddening significantly affects the fundamental parameters when using photometric 
techniques.

Finally, the equivalent widths of \ion{O}{i}T line do not show dependence to distances
derived by Vickers et al.\@(2015).

\section{Summary and conclusions}
\label{sec:summconc}

We presented a set of spectroscopic calibrations to obtain T$_\mathrm{eff}$, log\,$g$,
and [Fe/H] from equivalent widths of stellar spectra. The criteria choosen
for selection of the absorption features are similar to employed by MK classification system.
The equivalent widths for a total of 9 absorption features were measured.

We selected a total of 67 PAGB stars that include spectral types A and K, of which, 48 of them 
have a temperature between 6000 and 8000\,K (warm stars) and 19 have temperature from
4500 to 5500\,K (cold stars). For the determination of the spectroscopic calibrations we have 
identified the stellar parameters in the literature of 21 warm-PAGB stars and 5 cold-PAGB stars 
respectively.

We show the dependence of the stellar parameters with respect to the equivalent widths,
although the limitations present in the spectral ranges make it difficult to determine the
temperature, gravity and metallicity for all sample without previuos studies, i.e.
27 warm-and 14 cold-PAGB stars. These calibrations would be very useful to develop
suitable criteria for the rapid and accurate determination of fundamental parameters for
PAGB stars. The use of only spectral criteria is very important because it allows to define
the parameters for such objects, while the photometric indexes are often distorted by poor
known interstellar and circumstellar reddening. 

As future work it is possible to expand the spectral ranges and criteria in order to involve 
a great number of absorption features and to improve our spectroscopic calibrations for warm-and 
cold-PAGB stars using high-resolution spectra.

\section*{Acknowledgments}

We are grateful to Dr. Arturo Manchado for providing us the sample of low-resolution stellar 
spectral. We are thankful to Carolina Foundation for financial supporting to visit to Canarias 
Astrophysical Institute to Spain. We thank to Dr Sunetra Giridhar, Dr Armando Arellano Ferro and
Dr Valentina Klochkova for numerous comments and valuable sugestions on the text.
We express our gratitude to the anonymous referee for detailed comments that have improved the 
interpretation of the data and text.

\end{document}